\documentclass{Interspeech2024}
\usepackage{threeparttable}
\interspeechcameraready 
\usepackage{multirow}
\usepackage{hyperref}
\usepackage{authblk}
\title{CDSD: Chinese Dysarthria Speech Database}
\name[affiliation={1,2}]{Yan}{Wang$^{\#}$}
\name[affiliation={1,2}]{Mengyi}{Sun$^{\#}$}
\name[affiliation={3}]{Xinchen}{Kang}
\name[affiliation={1,2}]{Jingting}{Li}
\name[affiliation={4}]{Pengfei}{Guo}
\name[affiliation={5}]{Ming}{Gao}
\name[affiliation={1,2}]{Su-Jing}{Wang$^*$}

\address{
  $^1$CAS Key Laboratory of Behavioral Science, Institute of Psychology; $^2$Department of Psychology, University of the Chinese Academy of Sciences; $^3$Beijing Union University; $^4$ Jiangsu University of Science and Technology; $^5$University of Science and Technology of China 
  }
  
\email{wangsujing@psych.ac.cn}


\begin{document}

\maketitle

\begin{abstract}
    
   % Manuscripts submitted to Interspeech 2024 must use this document as both an instruction set and as a template. Do not use a past paper as a template. Always start from a fresh copy, and read it all before replacing the content with your own. The main changes with respect to previous instructions are highlighted in red.
    
  %  Before submitting, check that your manuscript conforms to this template. If it does not, it may be rejected. Do not be tempted to adjust the format! Instead, edit your content to fit the allowed space. The maximum number of manuscript pages is 5. The 5th page is reserved exclusively for acknowledgements and references, which may begin on an earlier page if there is space.
    
   % The abstract is limited to 1000 characters. The one in your manuscript and the one entered in the submission form must be identical. Avoid non-ASCII characters, symbols, maths, italics, etc as they may not display correctly in the abstract book. Do not use citations in the abstract: the abstract booklet will not include a bibliography.  Index terms appear immediately below the abstract. 

   %%%%%%%%%%%%%%%%%%%%%%%%%%%%%%%%%%%%%%%%由于他人难以理解构音障碍语音，构音障碍者在社会交流上存在困难。现在虽然ASR已经有广泛的应用，但是对于构音障碍者的识别仍存在巨大的挑战，这种挑战是由于构音障碍的数据匮乏导致的。为了丰富构音障碍的数据库，我们构建了CDSD，包括44人133小时的数据。据我们所知是目前最大的中文构音障碍数据库。我们也基于本数据库给出了一些benchmark，发现最好的情况下的CER达到16.4，同时我们也做了人类对构音障碍者语音的识别，其CER达到20.45，表明DSR是解决构音障碍者交流问题的有效途径。

Dysarthric speech poses significant challenges for individuals with dysarthria, impacting their ability to communicate socially. Despite the widespread use of Automatic Speech Recognition (ASR), accurately recognizing dysarthric speech remains a formidable task, largely due to the limited availability of dysarthric speech data. To address this gap, we developed the Chinese Dysarthria Speech Database (CDSD), the most extensive collection of Chinese dysarthria data to date, featuring 133 hours of recordings from 44 speakers. Our benchmarks reveal a best Character Error Rate (CER) of 16.4\%. Compared to the CER of 20.45\% from our additional human experiments, Dysarthric Speech Recognition (DSR) demonstrates its potential in significant improvement of communication for individuals with dysarthria. The CDSD database will be made  publicly available at \href{http://melab.psych.ac.cn/CDSD.html}{http://melab.psych.ac.cn/CDSD.html}.

%To further the development of speech recognition solutions for Chinese dysarthric speakers, this paper presents the CDSD: Chinese Dysarthic Speech Database. This database compiles speech and corresponding video recordings from 44 dysarthric speakers, 
%including 5 children, 
%totaling 133 hours of data. It stands as the largest Chinese dysarthria speech database to date. 
%The text pool of this database references materials from the AISHELL database and also provides text materials that are matched to the knowledge level of children with dysarthria. The database's collection environment is divided into laboratory and home settings, ensuring high-quality audio data while also reflecting the natural use of dysarthric speech in daily scenarios. 

%把下面这段改写成算法部分The database aims to improve the recognition accuracy of Chinese dysarthric speech, enabling Chinese patients with dysarthria to use speech assistant and voice input applications conveniently. 

%This article details the database collection process and the associated analysis results. The database will be made freely and publicly available at http://melab.psych.ac.cn/CDSD.html. 

%%%%%%%%%还需要把所有的模型所有的代码上传至GitHub，并把网站放上来。

\end{abstract}

\section{Introduction}
%%从交流表达到言语沟通
%言语对话是人与人之间交流的主要形式，良好的口语交流skills通常代表着更高的社交能力，我们生活在这个社会上离不开各种形式的言语交流。如果一个人失去了正常的言语交流能力，无法与他人流畅顺利的沟通交流，这不仅会影响到他的社会关系，还会对其心理健康造成负面影响。研究表明，患有言语障碍的青少年发生精神病学诊断和社会心理困难的风险增加（引文）。由此可见，让所有人听明白你说了什么，良好的语言表达能力对于人类的社会生活和心理健康来说是非常重要的。
Speech conversation is the fundamental to human communication~\cite{jowett1887politics,manson2013convergence}, and more proficient communication skills often reflect higher sociability~\cite{riggio1986assessment}. Dysarthria adversely affects individuals suffering from it by hampering their social relationships~\cite{braithwaite2017romantic} and negatively impacting their mental health~\cite{palmer2016does}. Studies have shown that dysarthria may place adolescents at increased risk for mental and physical health~\cite{rice1991social}. 
This suggests that fluent speech plays a pivotal role in fostering human social interactions and supporting psychological health.
\par
%构音障碍——种类、严重程度、病因
%然而，有相当一部分群体由于疾病或外伤等原因导致正常的言语功能受损。构音障碍属于运动型言语障碍，造成构音障碍的病因主要有脑瘫、帕金森病、肌萎缩侧索硬化症和脑卒中。按照病因不同，构音障碍患者可能会表现出不同程度的言语问题，例如说话语速更慢、发音不精确、不合理的停顿、清晰度下降等。这些症状导致构音障碍患者的言语难以被其他人理解，给他们增加了更多社交困难，也会对构音障碍者的心理健康造成影响。同时，构音障碍患者的言语不仅在社会交流方面会造成其他人理解困难，在人机交互方面，经典的语言识别算法对构音障碍言语的识别准确率也很低。
%A portion of the population experiences speech impairment due to factors such as diseases or injuries~\cite{ho1998speech}~\cite{whitehill2000speech}. 
Dysarthria, characterized by impaired speech due to motor deficits, is primarily caused by conditions such as cerebral palsy, Parkinson's disease~\cite{scott1983speech}, amyotrophic lateral sclerosis~\cite{makkonen2018speech}, and stroke~\cite{jerntorp1992stroke}. 
Depending on the etiologies cause, individuals with dysarthria may exhibit varying degrees of speech issues, such as slowed speech rate, imprecise pronunciation, irrational pauses, and decreased clarity~\cite{kent2000research,rowe2022characterizing}. 
%These symptoms make it challenging for others to understand the speech of individuals with dysarthria, leading to increased social difficulties and impacting their psychological well-being.
The speech of individuals with dysarthria poses challenges in their social communication and human-computer interaction (HCI). Especially, traditional automated speech recognition (ASR) systems often struggle to identify dysarthric speech.
\par
%构音障碍患者的ASR需求，提高独立性、社交能力等
%随着技术的发展，人与计算机的交互也越来越频繁，就像是人与人交流一样，言语对话也是人机的交互一个重要形式。通过语音识别，计算机可以理解用户的口头语音，并做出反应，这可以为用户提供更便利的操作方式。结合脑瘫等构音障碍的病因，许多构音障碍患者通常还面临着行动不便、需要他人陪护等问题，该群体正寻求一种简单、便利、多应用情景的人机交互形式，语音交互正好可以满足这一需求。如果计算机能够准确的识别构音障碍者的言语并做出交互，就不仅能更好的在日常生活中提供服务并降低陪护成本，还可以改善构音障碍者的社会交流和心理健康问题。然而，对于构音障碍患者的语音识别仍面临挑战，尽管现有的语音识别人机交互在识别健全人的语音时可以做到很低的准确率，但并不能很好的服务于构音障碍者。
%The interaction between humans and computers has become increasingly prevalent. Like interpersonal communication, speech is a crucial interaction between humans and computers. 
%自动语音识别有广泛的应用，包括智能家居、语音助手等。这些便利首先服务了正常人，而残疾人更需要这样的便利。然而，对于语音障碍的残疾人，却不能适用这样的便利。例如脑瘫，行动不便且构音障碍，
ASR offers a more convenient operation and has many applications, including smart home and voice assistants, primarily serving the needs of healthy individuals. Meantime, disabled people need such applications even more. Specifically, considering the etiologies of dysarthria, such as cerebral palsy, individuals with such disorders frequently encounter mobility challenges. They may require a more convenient form of interaction with computers~\cite{liu2021recent,shor19_interspeech}, with speech interaction serving as an ideal solution to this need.
\par
However, ASR for individuals with dysarthria still presents challenges~\cite{hu2023exploring,geng2022speaker}. Despite the high accuracy of ASR in interacting with healthy individuals, it serves individuals with dysarthria less effectively.
%ASR offers a more convenient operation and has many applications, including smart home and voice assistants. These applications serve healthy individuals first, and disabled people need such applications even more. Considering the etiologies of dysarthria, such as cerebral palsy, individuals with such disorders often face mobility challenges and may require a more convenient form of interaciton with computer~\cite{liu2021recent,wang2021improved,yue2022multi,geng2022speaker,tobin2022personalized,turrisi2022interpretable,hu2023exploring}, and speech interaction precisely meets this demand.

%Suppose computers can accurately recognize the speech of individuals with dysarthria and facilitate interaction. In that case, it enhances service provision, reduces caregiving costs in daily life, and addresses social communication and psychological health issues among those with dysarthria. However, ASR for individuals with dysarthria still presents challenges. Despite the high accuracy of ASR in interacting with healthy individuals, it serves individuals with dysarthria less effectively.
\par

%%%%%%%%%%%%%%%%算法相关工作（）

%\subsection{ASR for dysarthric speech}
%关于ASR算法
%ASR系统的主要功能是处理输入的音频信号，并生成与之对应的文本脚本。然而，ASR的性能通常受到训练数据的数量和质量的影响。
%ASR模型训练的识别准确度现状，经典语音模型与构音障碍语音模型的差别。
%用于识别构音障碍言语的ASR模型，即构音障碍语音识别DSR模型，可以更好的为构音障碍患者服务，提高他们的生活质量。但DSR的发展仍充满艰巨的挑战。首先，不同程度或不同病因的构音障碍语音的差异显著，需要自适应不同建模。其次，DSR模型性能的增强也需要大量数据的支持。在数据稀疏下的构音障碍语音识别还存在多特征多模态数据融合的问题。
The Dysarthric Speech Recognition (DSR) system, designed to recognize speech in individuals with dysarthria, can significantly enhance their quality of life. However, the development of DSR is still faced with formidable challenges. First, there are significant variability in dysarthric speech due to different degrees of severity or diverse etiologies, requiring adaptive modeling approaches. Second, enhancing the performance of the DSR system also demands extensive dysarthric speech data. Furthermore, in DSR under data sparsity, the challenge of multi-feature and multi-modal data fusion remains.
\par
Hernandez et al.~\cite{hernandez22_interspeech} demonstrated that speech representations pre-trained on extensive unlabeled data by models such as Wav2vec could significantly enhance ASR performance for dysarthric speech. 
Baskar et al.~\cite{baskar22b_interspeech} delved into integrating Wave2vec with either fMLLR features or x-vectors during the fine-tuning process. Meanwhile, Violeta et al. \cite{violeta22_interspeech} explored the efficacy of self-supervised learning frameworks, namely Wav2vec 2.0 and WavLM. Their comparative analysis revealed that the optimal Word Error Rate (WER) for severe dysarthric speech could reach 51.8\%. However, the significant disparity between normal and dysarthric speech  acts as a limiting factor for performance enhancement.
To bridge this gap, it is imperative to identify an effective approach. 
By pre-training on extensive speech corpora, neural networks can effectively learn prior knowledge from the training data. This strategy highlights the potential of utilizing large-scale speech data to refine representational models. Consequently, by fine-tuning based on specific data in downstream tasks, the capability of ASR systems to address the challenges presented by dysarthric speech can be significantly improved.
\par
In addition to acoustic features, vision constitutes another modality in human speech perception, indicating a bi-modal process~\cite{mcgurk1976hearing}. Furthermore, visual features remain unaffected by any degradation in acoustic signals and can thus provide valuable compensatory information for ASR systems.
In this context, the integration of visual information to enhance DSR performance, has been the subject of recent research efforts. For instance, the utilization of Bayesian gated control facilitates a strong integration of audio and visual modalities~\cite{liu2019exploiting}.
Furthermore, the multi-modal integration also help to address challenges such as severe voice quality degradation and the pronounced mismatch between dysarthric and normal speech patterns. For instance, Liu et al.~\cite{liu2020exploiting} proposed a novel strategy involving the generation of cross-domain visual features. Yu et al.~\cite{yu2023multi} developed a novel multi-stage fusion framework to improve the effectiveness of DSR systems.

%Their methodology encompasses a dual-phase fusion process, initiating with a visual fusion stage. In this phase, they meticulously extracted the facial speech functional areas frame-by-frame from the speaker's video and integrated these regions into a cohesive visual code. Subsequently, in the latter stage, this visual code was amalgamated with acoustic features leveraging the HuBERT framework to enrich the data representation. To optimize the AVSR model's performance, initial pre-training was conducted using a combined database of LRS2 and UASpeech, followed by a meticulous fine-tuning process. This refined AVSR model demonstrated remarkable performance improvements across various severity levels of dysarthric speech.

\par
%突出语音数据的低资源特点，需要用无监督或自监督算法。视觉辅助，结合视频视听识别，多模态识别可能更好。构音障碍的DSR模型与普通的ASR模型的差异，为什么DSR需要个性化。

%%%%%%%%%%%%%%%%%%%数据库相关工作
%\subsection{Existing database of Dysarthric Speech}
%关于已有的构音障碍数据库
%突出中文的少，带视频的少，普通语料库不能满足要求。
%现有的数据库规模，构音障碍DSR的识别仍有难度，尤其是中文构音障碍语音数据库的确实。

%由此可见，提升针对构音障碍患者的自动语音识别正确率，需要依靠对大量的构音障碍者的语音数据进行训练。然而，基于构音障碍者说话困难的根本原因，构音障碍者的语音数据筹集过程充满了挑战。并且，构音障碍者的语音数据标注也十分困难，进一步加大了构建构音障碍数据库的难度。基于这些原因，相比起正常语音数据库来说，构音障碍数据库相对较少。虽然现在已有一些数据库可以支持DSR的训练，但这些数据库相当一部分属于英文语音数据，并且数据的模态和应用场景较单一。
Compared to the typical speech databases, the sizes of dysarthric speech databases are pretty small. However, individuals with dysarthria have difficulty speaking, so collecting speech data from individuals with dysarthria is challenging. Moreover, annotating speech data from individuals with dysarthria is particularly challenging, further increasing the difficulty of constructing databases for dysarthric speech. Therefore, dysarthric speech databases are relatively scarce compared to typical speech databases.
\par
The constructed databases for disordered speech, such as Whitaker~\cite{deller1993whitaker}, UA-Speech~\cite{kim2008dysarthric}, Torgo~\cite{rudzicz2012torgo}, EasyCall~\cite{turrisi21_interspeech}, and Euphonia corpora~\cite{macdonald2021disordered}, are predominantly English-speaking disordered speech databases. 
% They primarily consist of audio data, needing more multimodal data that includes corresponding video signals, which is not conducive to Chinese DSR training.
%%%%前人各种其他语言的数据库的工作，例如Whitaker、UA-Speech、Torgo、Euphonia语料库
%早期的构音障碍语音数据库，包括起源于1993年的Whitaker项目，由密歇根州立大学和东北大学合作完成。在最初阶段，Whitaker并未专门设计用于神经网络的训练，因此数据量不足以支持基于神经网络的语音识别模型的训练。
%EasyCall通过创建与常见命令词相似的语义或语音上的词汇和句子，来解决构音障碍用户发音偏离的挑战。这种方法不仅提高了识别系统的准确性，而且为开发语音识别技术提供了重要资源，以支持构音障碍者通过辅助技术获得更好的语音识别体验。
%目前最广泛使用的构音障碍语音数据库包括UA-Speech和Torgo。这两个数据库都提供了语音清晰度评级，有助于在训练后评估说话者的语音清晰度。这种基于语音清晰度的分类为构音障碍在语音识别任务中的严重程度分类奠定了基础。
%鉴于中文和英文的发音方式、语言形态、句子结构上存在显著差异，并且中文比起英文由更多的同音字和多音字，导致现有的英语构音障碍数据库不能满足中文DSR任务的特殊要求（引用文献）。 由此可见，为推动当前中文DSR的发展，急需足够的中文构音障碍语音数据以供训练。 已有的中文构音障碍数据库包括广东话构音障碍语音语料库（CUDYS）和普通话亚急性卒中构音障碍多模态（MDM）数据库，但这些数据库均存在局限性，例如样本量少、不适用于DSR训练等诸多问题。 为填补这一资源领域上的空缺，我们致力于开发一个适用于DSR训练的关于脑瘫患者的中国构音障碍语音数据库。
The distinct pronunciation, morphology, and syntax of Chinese, along with its many homophones and polyphones, mean that English dysarthric speech databases fall short for Chinese DSR tasks. Clearly, advancing Chinese DSR urgently requires a substantial Chinese dysarthria speech dataset for training.
\par
Existing Chinese dysarthria speech databases include the Cantonese Dysarthric Speech Corpus~\cite{wong2015development} and the Mandarin Subacute Stroke Dysarthric Multimodal database~\cite{liu2023audio}. However, the size of these existing Chinese dysarthric speech databases does not exceed 10 hours, which is relatively small. To address these resource gaps, we are committed to developing a Chinese dysarthric speech database for DSR training, focusing on individuals with cerebral palsy.

%在已构建的中文构音障碍语音数据库中，迄今为止仅报告了两个数据库：广东话构音障碍语音语料库（CUDYS）[23]和普通话亚急性卒中构音障碍多模态（MDM）数据库[24]。CUDYS主要研究构音障碍语音发音和韵律的声学特征，包括语速、语调和强度。另一方面，MSDM专注于亚急性中风患者作为参与者，使用专业级麦克风捕获音频数据，并在演讲过程中与摄像机同步记录面部运动数据。这些资源为客观地量化和识别言语产生中的病理差异提供了宝贵的支持。然而，这两个数据库都有相对较小的数据集，每个数据集包含不到10小时的构音障碍语音，这使得它们不足以训练全面的语音识别模型。（上述构音障碍语音数据库的具体信息见表，做表）
%In the existing Chinese Dysarthric Speech Databases, only two have been reported so far: the Cantonese Dysarthric Speech Corpus (CUDYS)~\cite{wong2015development} and the Mandarin Subacute Stroke Dysarthric Multimodal (MDM)~\cite{liu2023audio} Database. The research conducted on CUDYS primarily focuses on the acoustic features of pronunciation and rhythm in dysarthric speech, including speech rate, intonation, and intensity. On the other hand, MDM is centered around speakers with subacute strokes, utilizing professional-grade microphones to capture audio data and synchronously recording facial movement data with cameras during speech. These resources provide valuable support for objectively quantifying and identifying pathological differences in speech production. However, both databases have relatively small database, with each containing less than 10 hours of dysarthric speech, making them insufficient for training comprehensive speech recognition models.
%%可以表现我们CDSD与其他中文构音障碍数据库的区别，有发音器官的运动数据，CDSD有面部运动的视频

%\subsection{Contribution}
\par
%本文结构或贡献
%我们的数据库不仅是为数不多的中国构音障碍语音数据库之一，同时也是现有的规模最大的中文构音障碍数据库。我们的数据库总时长为134小时，代表了更大规模的资源，为汉语语境的困难语音研究提供了更广泛的样本。同时CDSD数据库中的大部分数据是参与者在家中使用移动设备记录的，从而创造了一个更自然的环境。这样的录制环境更接近于构音障碍用户对于DSR的日常使用情景，从而提高了数据的生态效度并提高了语音识别模型的鲁棒性。最后，在DSR算法方面，我们引入说话人依赖语音识别的概念，为解决构音障碍说话者之间存在显著差异的语音识别挑战提供了一个有前途的解决方案。
To our knowledge, the Chinese Dysarthria Speech Database (CDSD) is not only among the rare Chinese dysarthria speech databases available but also stands as the largest in scale compared to any existing databases of its kind. With a total duration of 133 hours, CDSD represents a more extensive resource, offering a broader sample for the study of challenging speech in the Chinese language context. 
%2、通过人和计算机dsr的比较实验，证明了DSR在改善构音障碍者社会交流方面有一定的潜力。
Furthermore, through comparative experiments between human and computer, DSR has shown the potential to improve social communication of individual with dysarthria. 
%Notably, the majority of the data in the CDSD is recorded by speakers using smartphone in their homes.
%Such recording conditions closely resemble everyday usage scenarios of DSR, thereby enhancing the ecological validity of the data and improving the robustness of speech recognition models.
Finally, we conducted a preliminary exploration into the optimal data collection length for speaker-dependent DSR by comparing training sets of various sizes.
%In addition, regarding DSR algorithms, we introduce the concept of speaker-dependent speech recognition, providing a promising solution to address the speech recognition challenges due to significant variations among speakers with dysarthria.
%1、数据库规模，是目前最大的中文构音障碍数据库

%3、对构音障碍者的低资源语音识别做了初步的探讨,通过对比不同尺寸的训练集，对构音障碍数据库的最合理的数据收集长度进行了初步的探索

%%%%%%%%%%%%%%%%%%%%%%%%%%%%%%%%%%%%%%%%%%%%%%%%%%%%%%%%%%%%%%%%%%%%%%%%%%%%%%%%%%%%%%%%%%%%%%%%%%%%%%%%%%%%%%%%%%%%%%%%%%%%%%%%%%%%%%%%%%%%%%%%%%%%%%%

%%%%%%%%%%%%%%%%%%%%%%%%%%%%%%%%%%%%%%%%%%%%%%%%%%%%%%%%%%%%%%%%%%%%%%%%%%%%%%%%%%%%%%%%%%%%%%%%%%%%%%%%%%%%%%%%%%%%%%%%%%%%%%%%%%%%%%%%%%%%%%%%%%%%%%%

%\section{Related Work}

%%%%%%%%%%%%%%%%%%%%%%%%%%%%%%%%%%%%%%%%%%%%%%%%%%%%%%%%%%%%%%%%%%%%%%%%%%%%%%%%%%%%%%%%%%%%%%%%%%%%%%%%%%%%%%%%%%%%%%%%%%%%%%%%%%%%%%%%%%%%%%%%%%%%%%%

%%%%%%%%%%%%%%%%%%%%%%%%%%%%%%%%%%%%%%%%%%%%%%%%%%%%%%%%%%%%%%%%%%%%%%%%%%%%%%%%%%%%%%%%%%%%%%%%%%%%%%%%%%%%%%%%%%%%%%%%%%%%%%%%%%%%%%%%%%%%%%%%%%%%%%%

\section{Database construction}

%\subsection{Description of the database}
%3.1 Description of the database
%数据类型与数据规模
%CDSD数据库中包括由手机或专业录音麦克风等不同设备采集的构音障碍音频数据，是在居家环境以及实验室环境下录制的，同时还包括一部分与音频相对应的视频数据。CDSD是现在已知最大的可用于神经网络训练的多模态中文构音障碍数据库。   

%目前，我们已经收集了一共44名参与者的语音数据，我们总共收集了134小时的样本于部分视频数据。CDSD的所有数据一共分为两个部分。A部分的音频数据来自于44个参与者，每位参与者录制的1个小时的语音数据，其中的9位参与者还提供了与他们录音对应的视频数据。B部分的音频数据为已有参与者中的9位，额外录制的10小时语音音频数据。同时，我们还在匿名化之后保存了构音障碍参与者疾病类型、录制环境、年龄、口音等说话人信息，可以作为模型训练的辅助特征。(做表展示PART A和PART B)
%这8名参与者是来自PartA的
%Part A中的44名参与者有8名每人额外录制了10h构成了Part B
The CDSD database has collected speech data from 44 speakers, including 124 hours of audio data and 9 hours video data. %These data were collected in two different environments using separate devices. 
Based on the two different durations of each subject's recording, we divided the database into two parts.
% To accommodate the specific data needs of speaker-dependent and subject-independent recognition tasks,
Specifically, Part A includes 44 hours of audio data from all 44 speakers, 1 hour of recording each. Additionally, it includes video recordings that are synchronized with the audio from the nine speakers. Part B consists of 80 hours of audio data, 10 hours recorded by each of the 8 speakers in Part A. The detail of CDSD is listed in Table~\ref{table1}. 
%Part B was recorded among 8 speakers from the 44 speakers in Part A, and includes 80 hours of audio data, with each recording lasting 10 hours

%Part B included an additional 80h of audio data from eight speakers originally involved in Part A.

%%Part B were recorded by 8 speakers amoung all 44 speakers in Part A, includes 80 hours of audio data, each recording 10 hours.

%%%%%%%%%%%%%%%%%%%%%%%%%%需要重新画表

%\begin{table}[h]
%\caption{All data in the CDSD database are divided into two parts, where A and V are the audio and video data respectively, and E represents the data size for each speaker. Speakers in Part B recorded Part A as well. }
%\centering
%\begin{tabular}{ccccc}
% \toprule
 %\centering
%\textbf{CDSD} & \textbf{Data} & \textbf{speakers} &\textbf{E} & \textbf{Size} \\  \midrule
%\raisebox{-1.5ex}[0pt]{Part A} & A & 44 & 1h & 44h \\
%       & V & 9 & 1h & 9h \\ \midrule
%Part B & A & 8 & 10h & 80h \\ \midrule
%ALL    &       & 44 &  & 133h \\
% \bottomrule
%\end{tabular}
%
%\label{table1}
%\end{table}

\begin{table}[h]
\caption{CDSD overview. \# indicates the amount of the speakers, and D/P represents the recording duration per speaker. *The speakers in Part B were also involved in Part A. }

\centering
\begin{tabular}{ccccc}
 \toprule
 \centering
\textbf{CDSD} & \textbf{\#} &\textbf{D/P} & \textbf{Total audio} & \textbf{Total video} \\  \midrule
Part A & 44  & 1 h & 44 h & 9 h \\
Part B &   8\textsuperscript{*}  & 10 h & 80 h & \textbackslash{}     \\
All   &    44 & \textbackslash{} & 124 h & 9 h \\
 \bottomrule
\end{tabular}

\label{table1}
\end{table}

%B部分的8名被试可以加标注，即pb中的8名参与者是

\subsection{Data collection}

%3.2.1 被试  来源 公益团队支持、知情同意、隐私保护、其他人文关怀

\textbf{Participants}: In the preparatory phase of data collection, each speaker signed informed consent prior to the recording. A total of 44 speakers were recruited, 39 speakers over 18 years old, and the remaining 5 speakers were younger than 18. For those minor speakers, parental or guardian consent was obtained, along with the minors' assent. Additionally, the 9 speakers recorded on video were provided with additional informed consent forms. Speakers were informed of their right to discontinue participation at any point without consequence.

\par
To protect speakers' privacy, speakers' names were not collected in CDSD. Instead, after speech data collection, each data was assigned a speaker representative serial number. In addition, we have collected information about the speakers' etiologies of dysarthria, recording environments, ages, accents, and other speaker-related details. The overall information of the speakers in the CDSD is shown in Table~\ref{table2} .
\par

%\begin{table}[htbp]
%\centering
%\caption{Group Statistics}
%\label{table2}
%\begin{tabular}{@{}lllr@{}}
%\toprule
%\multicolumn{3}{l}{\textbf{Factor}} & \multicolumn{1}{r}{\textbf{Overall}} \\
%\multicolumn{3}{l}{\hspace*{1em}Categorization} & \multicolumn{1}{r}{(N=44)} \\
%\midrule
%\textbf{\hspace*{1em}Sex} & & & \\
%\hspace*{2em}Female & & & 18 \\
%\hspace*{2em}Male & & & 26\\
%\textbf{\hspace*{1em}Age} & & & \\
%\hspace*{2em}Adults & & & 39 \\
%\hspace*{2em}Children & & & 5  \\
%\textbf{\hspace*{1em}Recording devices} & & & \\
%\hspace*{2em}Smartphone & & & 39 \\
%\hspace*{2em}ZOOM F8n & & & 5  \\
%\textbf{\hspace*{1em}Etiology} & & & \\
%\hspace*{2em}Cerebral Palsy & & & 33 \\
%\hspace*{2em}Other Diseases & & & 11  \\
% \bottomrule
%\end{tabular}
%\end{table}

\begin{table}[htbp]
\centering
\caption{The overall information of the speakers.}
\label{table2}
\begin{tabular}{ccr}
\toprule
\multirow{2}{*}{\textbf{Factor}} & \multirow{2}{*}{\textbf{Categorization}} & \textbf{Overall} \\
& & (N=44) \\
\midrule
\multirow{2}{*}{\textbf{Sex}} & Female & 18 \\
                      & Male & 26 \\[0.35\normalbaselineskip]
\multirow{2}{*}{\textbf{Age}} & Adults & 39 \\
                      & Children & 5 \\[0.35\normalbaselineskip]
\multirow{2}{*}{\textbf{Etiology}} & Cerebral Palsy & 33 \\
                      & Other Disease & 11 \\[0.35\normalbaselineskip]
\multirow{2}{*}{\textbf{Recording devices}} & Smartphone & 39 \\
                      & ZOOM F8n & 5 \\

\bottomrule
\end{tabular}
\end{table}

%不能用“预实验”来表示“在正式收数据前的流程验证实验”，直接说过程和目的就行
Notably, the speaker cohort for this database includes two authors of this article, both diagnosed with dysarthria. Their dual roles as researchers and speakers enriched the study with nuanced insights, fostering a research environment characterized by inclusivity. For example, prior to the formal recording, we asked one of the authors to record for 10 hours as the first speaker. He/She finds it challenging to speak continuously for long duration, as it may affect pronunciation accuracy. Therefore, speakers were allowed to submit recordings in segments based on their individual conditions. Furthermore, sentences reported by the first speaker as challenging to articulate were excluded from subsequent data collection phases.
%the first speaker 王老师，

%3.2.3 文本材料
%Symbols such as $<$, $>$, $[$, $]$, $\sim$, $/$, \textbackslash, =, etc., are removed to facilitate subsequent annotation.  
\textbf{Text pool}: The text pool for constructing the CDSD consists of two types of texts. In particular, the first type is sourced from the AISHELL-1  database~\cite{aishell1}. This text pool cover various domains of language diversity, providing a better representation of daily speech usage. Additionally, it encompass a wide range of commonly used Chinese words and characters, enhancing the database's universality and improving model robustness. Furthermore, texts are purged of inappropriate content related to sensitive political issues, user privacy, pornography, violence, etc.
Meantime, the second type comprises elementary and middle school speeches and fairy tales. This is because that a small portion of our speakers are children. To accommodate children's reading habits and match children's level of literacy knowledge, we extracted speeches and fairy tales from the internet as the second type of text.

%录制设备    多种，包括手机、录音声卡、不同设备保存的格式统一 WAV、不同的设备用的算法是一样的、视频通过手机录制、为什么要录视频、通过多模态来增加识别率（视频与语音一起识别可以增加语音的识别率的引文）
%%因为参与录制的参与者大部分是脑瘫患者，他们在患有构音障碍的同时通常还伴有运动障碍，这就导致让参与者集中进行录制音频是十分困难的。因此，我们采取两种不同的环境和硬件进行音频录制。一些演讲者是在大约 10 平方米的录音室中使用专业录音设备——ZOOM F8n 现场录音机录制的。ZOOM F8n 具有高质量的前置放大器模拟输入信道，可产生更高的音频录制质量。这种高质量的记录允许在以后的数据分析中最大限度地减少潜在的干扰因素。同时，我们为行动不便的参与者采用了分布式录音方法，让他们可以在安静的家庭环境中舒适地使用移动设备录制自己的音频。这种录音方法可以很好的模拟说话者的日常语音识别任务，增强音频数据的生态效度并增强语音识别模型的鲁棒性。 同时，我们还通过手机录像采集了部分参与者的视频数据，与其音频数据对应，以此获得多模态的数据来提高语音识别准确率。
%所有提到的硬件设备在正式录制之前都经过了严格的测试，以确保其效率并符合所需的录音标准。这个过程保证了我们可以高质量地收集志愿者的音频样本，而不会出现任何明显的损失或失真。不同场景和设备录制的音频数据均以WAV格式保存，并用于同一个算法进行识别。
%Since the most of speakers are cerebral palsy sufferers, who often experience motor impairments alongside dysarthria, considering the health status of speakers, it is challenging to engage them in prolonged audio recording tasks. Therefore, we adopted two different environments and hardware setups for audio recording.
\par
\textbf{Recording devices}: Two types of recording devices were used to collect data: the ZOOM F8n field recorder and smartphone. Specifically, some speakers recorded their audio in a recording studio of approximately 10 square meters using professional recording equipment — the ZOOM F8n field recorder. The ZOOM F8n has high-quality preamplifier analog input channels, producing superior audio recording quality. This high-quality recording allows for the minimization of potential interference factors in subsequent data analysis. Meanwhile, to accommodate speakers with motor impairments, speakers could record their audio comfortably using Smartphones within the tranquil confines of their home environment. Using smartphone recording effectively simulates the speakers' daily speech recognition tasks, enhancing the ecological validity of the audio data and improving the robustness of the speech recognition model. Additionally, video data of some speakers were recorded using smartphone, synchronized with their corresponding audio recordings. The acquisition of multimodal data could facilitate the improvement of speech recognition accuracy.
\par
All recording devices underwent rigorous testing before formal recording to ensure efficiency and compliance with the required recording standards. This process ensured the audio data collection with high quality and without any noticeable loss or distortion. Additionally, audio data recorded in different scenarios and with different devices were stored in WAV format and analyzed using the same recognition algorithm.

\par
\textbf{Recording process}: First, the quietude of the recording environment was imperative, regardless of whether the setting is a recording studio or a home.
Then, during the audio and video recording processes, speakers were instructed to ensure that the microphone was approximately 20-40 centimeters away from their mouth. Additionally, they were required to maintain stable recording equipment and consistent volume levels throughout the recording process. Throughout the video recording process, speakers were instructed to maintain their facial images centrally aligned on the screen, ensuring both shoulders and the full movement of their lips were clearly visible.

\subsection{Data annotation}
%3.3 Data preprocessing and Annotation

Due to the substantial differences between the speech of individuals with dysarthria and healthy individuals, as well as the differences in the speech of each individual with dysarthria, annotating the speech of dysarthric speakers posed particular challenges. Speakers might have made reading errors or skipped words when reading text, requiring annotators to confirm the speech's accuracy repeatedly. Additionally, some speakers had severe dysarthria, resulting in very unclear pronunciation. And some other speakers had noticeable regional accents. Both issues made speech recognition more complicated for annotators.

\par

 %Audio data with adequate quality were imported into the iFlytek AIBIAOKE annotation platform and transcribed verbatim by the original audio content to ensure consistency between the annotated and spoken audio. Each marked voice waveform was preceded and followed by a 0.1-second buffer to prevent any "clipping" phenomenon. The audio editing work is done by five proficient annotators skilled in using the AIBIAOKE annotation platform. Additionally, an inspector reviews the quality of all edited audio. Before the editing process, all annotators receive standardized training and follow a unified set of editing standards and procedures.

The audio annotation task was conducted by five proficient annotators using the AIBIAOKE annotation platform\footnote{\href{http://124.243.239.193:8081/\#/login}{http://124.243.239.193:8081/\#/login}}. Prior to annotation, all annotators underwent standardized training and adhered to uniform editing standards and procedures. Then,  annotators reviewed the quality of all audio data and contacted the speaker for re-recording if necessary. After the audio data passed inspection, it was imported into the AIBIAOKE annotation platform. Annotators transcribed the text verbatim based on the original audio content to ensure consistency between the audio and the text. A 0.1-second buffer was placed before and after each marked speech waveform to prevent any  ``clipping" phenomenon.
Non-speech segments of human voice lasting $\geq$ 0.5 seconds, such as static noise, laughter, breathing, coughing, and singing, were marked as NOISE.

\section{Experiments}
Experiments are conducted on Part A and Part B of CDSD. In particular, Part A comprises 44 speakers, with each contributing roughly one hour of data, amounting to a total of approximately 44 hours. The data of Speaker \#2 in Part B was removed from the experiment because it was incorrectly annotated. The correct version has now been updated to the published database. Therefore, Part B remains 7 speakers for performance comparison, with each contributing roughly ten hours of data, amounting to a total of approximately 70 hours. The data is segmented into training, development, and test sets in an 8:1:1 ratio.
\par 
Two kinds of feature: Fbank and Wav2vec, are utilized in our experiments. The Fbank serves as a traditional spectrum feature, while Wav2vec, a self-supervised learning representation, significantly enhances DSR tasks~\cite{hernandez22_interspeech, baskar22b_interspeech, violeta22_interspeech}. The Wav2vec is modeled on WenetSpeech\footnote{\href{https://github.com/TencentGameMate/chinese_speech_pretrain}{https://github.com/TencentGameMate/chinese\_speech\_pretrain}}.
\par
A end-to-end model is trained as the baseline model based on the ESPnet toolkit. Specifically, Conformer~\cite{gulati20_interspeech} is used as the ASR config and RNN is used as the inference config. 
We adopted two training approaches for our models. Initially, we directly train models on Part A or B of CDSD. Separately, we first train models on the AISHELL-1~\cite{aishell1}, AISHELL-2~\cite{du2018aishell}, and WenetSpeech~\cite{zhang2022wenetspeech} databases to obtain pre-trained models respectively, which we then fine-tune on CDSD Part A or B.
Table~\ref{tab:result} lists character error rates (CERs).
Due to computational resource limitations, the model was not trained on WenetSpeech database using Wav2wec features.
\begin{table*}[h!]
  \caption{CERs with Fbank and Wav2vec on various pre-trained models. [D/T] D and T mean the CER on the development set and the CER on test set. The arrow ``$\rightarrow$" indicates that pre-training is conducted on the databases of the former, followed by fine-tuning on the Part A or Part B of the CDSD database.}
  \label{tab:result}
  \centering
  \begin{tabular}{cccccc}
 \toprule
 \centering
\textbf{ }& \textbf{Features}& \textbf{Part X}& \textbf{AISHELL-1$\rightarrow$Part X}& \textbf{AISHELL-2$\rightarrow$Part X}& \textbf{ WenetSpeech$\rightarrow$Part X}\\
 \midrule
\raisebox{-1.5ex}[0pt]{\textbf{Part A}}&Fbank& 24.9 / 24.9 & 20.9 / 21.2  &20.5 / 20.7 &16.5 / 16.4 \\
&Wav2vec&22.0 / 22.2 & 25.6 / 25.6  &26.9 / 27.1 & No Answer\\
\\
\raisebox{-1.5ex}[0pt]{\textbf{Part B}}&Fbank &31.9 / 30.2& 28.4 / 26.8& 26.3 / 24.7 &23.7 / 22.2\\
&Wav2vec &30.4 / 28.7 &30.8 / 29.2 & 30.5 / 29.0 & No Answer\\

 \bottomrule
\end{tabular}
\end{table*}
\par

\begin{table*}[h!]
  \caption{CERs of computer and human transcription of each speaker in Part B.}
  \label{tab:result2}
  \centering
  \begin{tabular}{ccccccccc}
 \toprule
 \centering
\textbf{Speaker ID}& \textbf{\#1}& \textbf{\#4}& \textbf{\#6}& \textbf{\#8}& \textbf{\#9}& \textbf{\#12}& \textbf{\#20}\\
 \midrule
Computer &11.9 / 11.6&29.3 / 27.0 & 37.9 / 37.1 & 21.7 / 19.4 & 20.1 / 19.8 & 11.1 / 10.5 & 8.5 / 9.4 \\
Human& 6.2 & 10.8 & 60.5 & 48.9 & 28.2 & 12.8 & 15.5\\
 \bottomrule

\end{tabular}

\end{table*}
\par

The scales of Part A / B of CDSD, AISHELL-1, AISHELL-2, and WenetSpeech are 40+ / 80
+, 100+, 1000+ and 10000+ hours, respectively. As the data scale increases (referring to the pre-training), the CERs of using Fbank as features decreases, and shows a significant reduction with pretraining on WenetSpeech.
%, because the scale of WenetSpeech data is large, and its pre-trained model can achieve superior performance. 
%\par
Meanwhile, the Wav2vec feature has better performance on the case of modelling directly on the CDSD without any pre-trained models. This proves that Wav2vec can effectively represent the strong variability of dysarthric speech. However, this performance did not surpass that achieved using the Fbank acoustic features in AISHELL-1 and AISHELL-2, attributing the limitation to the significant disparity between normal and dysarthric speech.

%%%%%%%%%%%%%人机对比实验%%%%%%%%%%需要加入表格
We designed an experiment comparing human and computer on dysarthric speech recognition. The experiment included 10 participants, divided into two groups of five each based on whether they had experience communicating with individuals with dysarthric speech. All participants were asked to recognize and transcribe 17 dysarthric speech utterances within a 10-minute time limit. All utterances were randomly selected from the CDSD Part A, were approximately 4 seconds in length, and were ensured to be clear and unambiguous. 
\par
The mean CER of the participants with experience in communicating with dysarthric speakers was 20.45$\%$, and the inexperienced participants were 35.71$\%$. Such comparisons show that sufficient experience is required to understand dysarthric speech accurately. Meantime, compared with the best result in Table~\ref{tab:result}, the computer's speech recognition ability is superior to that of humans, highlighting the potential of DSR to enhance social interaction for individuals with dysarthria.
%%%%%%%%%%%%%%%%%%%%%%%%%%%%%%%%%%%%%%%%%%%%%%%%%%%%%%%%%%%%%%%%%%%%%%%%%%%%%%%%%%%%%%%%%%%%%%%%%%%%%%%%%%%%%%%%%%%%%%%%%%%%%%%%%%%%%%%%%%%%%%%%%%%%%%%
%在此基础上，我们又让语音识别模型与有构音障碍者交流经验的被试进行更多的语音识别与转录上的对比。我们将CDSD数据库Part B的语音投入说话人相关的模型训练，采用Fbank特征，并分析出以Part B中每一位speaker作为development set和test set的CER。同时，对Part B中的每个speaker随机抽取10个4秒左右的utterances，让被试对从PartB的每个speakers中抽取的语音进行识别转录。对比结果如表所示。

%%%首先，对part b每个speaker随机抽取10个utterances，一共得到80个utterances。其次，让参与者识别与转录这些utterances，并计算出所有参与者对每个speak的语音识别结果的平均CER。

%解释自动的结果
% 在Table 3中，尽管Part B相较于part A 有更高的数据时长，但是part A 有更多的被试数目(80h/8 vs 44h/44)。可以看出在被试量有限的情况下，模型的泛化能力较差，无法学习到更多类型的语音特征，所以part A的语音平均识别性能优于part B. 这让我们好奇，在单个被试语音环境下的模型性能。因此，我们对Part B中的样本进行了speaker-dependent的测试。
In Table~\ref{tab:result}, superior average speech recognition performance is demonstrated on Part A compared to Part B. Despite Part B having a longer data duration compared to Part A, Part A has a higher number of subjects (70 hours / 7 vs. 44 hours / 44). This indicates that under the limitation of subject quantity, the model exhibits poorer generalization ability, unable to learn a more diverse set of speech characteristics. This piqued our curiosity about the model's performance in a single-subject speech environment. Consequently, we put Part B into speaker-dependent model training, using Fbank features, and calculated CER with each speaker's sample in Part B as the development set and test set. The experimental results are shown in Table~\ref{tab:result2}.
\par
Besides, we compared the recognition result of computer with that of participants who had experience communicating with individuals with dysarthria. We randomly selected 10 utterances for each speaker in Part B. The participants were asked to recognize these utterances and their CERs were calculated.

%%%%%%%%%%%%
%按照严重程度划分的人机识别准确率对比，可以在期刊论文里细化
%%%666%人类语音发音主要通过两个阶段，声带和口腔。可以从视频上看到，过于严重的构音障碍者，口腔部分的肌肉动作很少，导致发音不完善，声音变化也相对少，人类识别者难以识别这种微小的声学变化，导致人类对这类严重的构音障碍患者的语音识别CER较高。而语音识别模型经过特征提取，可以识别这些更微小的变化，所以对6号的识别CER比人类识别低。
%%%%124%每个人是怎么听怎么识别的，识别语音的思考过程。回访5个识别人，总结人手工识别的思维过程。
%%%%%%%%%%%%此处再加上实验过程？可能与上文重复

%2号的语音样本比其他构音障碍样本更独特（区别）：区别于多数构音障碍者的语音气息不足的病症，2号speaker在说话时气息稳定且声音足够响亮，并且没有很大的声调偏差，其构音障碍主要存在于口腔动作的障碍，而非声带区域的disorder。这使得2号speaker在某些口腔动作依赖较少的发音上更接近正常人的语音。Furthermore，目前用于训练的样本量仍然较少，这就容易模型导致在对2号speaker语音进行识别的过程中，预训练模型本身发挥的作用更大，样本的影响较小，从而导致2号的发音很容易被误判成其他字，造成假阳，预训练的影响可能会高于样本训练对其的影响，造成高CER的产生。

%表4中超过半数的对比结果都表明计算机的DSR好于人类。（但构音障碍者出现较大语音差异时，计算机仍存在不能准确识别的问题。）但是每个构音障碍者存在不同的声学特征，这会导致计算机在识别不同的构音障碍者的语音时出现较大的DSR性能差异。However, each dysarthric speaker has various acoustic features(characteristics), which made the DSR performance of the computer vary significantly when recognising the speech of different dysarthric people.

%注：part B中的一个speaker数据存在标注错误，故在实验中删除，正确版本现已更新至发布的数据库中。
%\footnote{The data of Speaker \#2 in Part B was deleted from the experiment because it was incorrectly annotated. The correct version has now been updated to the published database.\label{3}}
%One of the speaker data in part B was deleted from the experiment because it was incorrectly annotated, and the correct version has now been updated to the published database.

More than half of the comparisons in Table~\ref{tab:result2} show that the computer's DSR is better than the human's. However, each dysarthric speaker has various acoustic features, which makes the DSR performance of the computer exhibits fluctuations. % For instance, the speech sample of the speaker \#2 is distinct from other dysarthric speech samples, exhibiting stable breath and a clarity that approaches that of typical speech. 
Compared to unimpaired speech, there are longer pauses in their expression and a lack of coherence between syllables. This characteristic has led to challenges in model training due to a limited database, possibly causing an over-reliance on pre-trained features. Hence, the model struggles to construct contextual relationships from clear speech punctuated by atypical pauses, often misclassifying the pronunciations from this speaker and resulting in a high CER.

\par
%%%%%%%%%%%实验过程：最高和最低CER的speaker在不同训练量上再进行CER的比较；2号和20号，完整的10小时，一半的5小时，四分之一的2.5小时，三种数据训练量进行对比。
% 为了研究采集样本音频时长是否对subject-depent DSR是否足够，我们在表四中，选择最高和最低CER的speaker在不同训练上（等比递减）进行DSR性能比较，即8h, 4h,2h和1h。与此同时，development set和test set保持不变，仍为1h。如图所示，当训练集数据时长大于等于4小时的时候，模型性能的提升非常平缓。由此，可以得出结论，
%%%%在一个大规模的数据集上，比如wenetspeech的预训练模型上，对构音障碍者的语音识别进行微调和训练的时候，每个构音障碍者需要的有标注的数据大概在5个小时左右
%这对之后的构音障碍语音数据的收集有指导意义，特别是在构音障碍语音的采集和标注都非常困难的清情况下避免数据库构建的成本升高。
\begin{figure}[t]
  \centering
  \includegraphics[width=\linewidth]{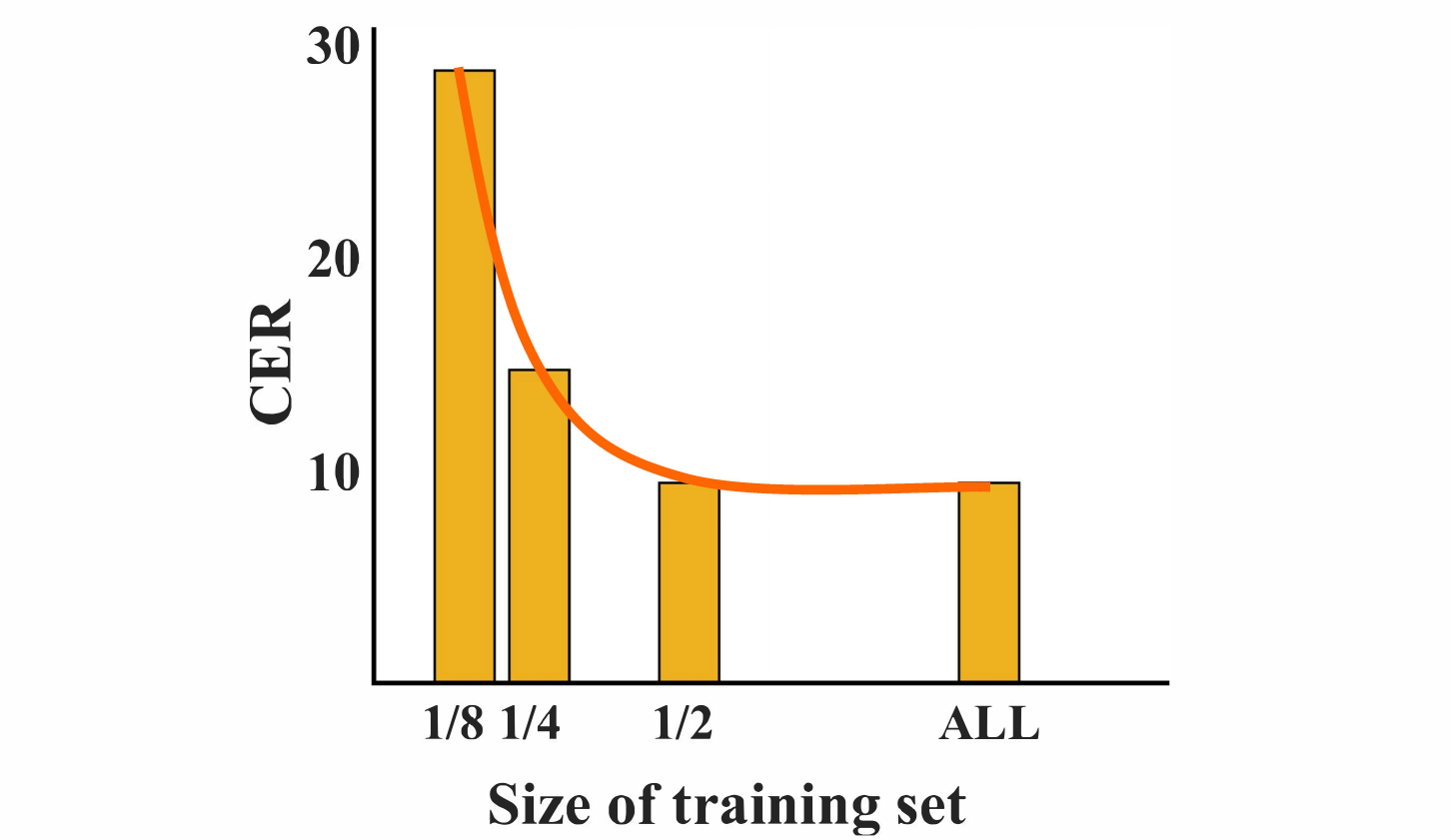}
  \caption{Optimal training data quantity determination.}
  \label{fig:diffduration}
\end{figure}
%Constructing speaker-dependent models requires a substantial amount of training data, and the collection and annotation of dysarthric speech data are relatively challenging, resulting in a limited availability of dysarthric speech data. 
Low-resource speech recognition aims to achieve optimal performance using minimal data. To determine the optimal training data quantity for speaker-dependent DSR performance, we compared the DSR performance of the speaker with the lowest CER listed in Table~\ref{tab:result2} across varying training durations (in a proportional decrement). Meanwhile, the durations of the development and test sets remained constant at 1 hour. As illustrated in Fig.~\ref{fig:diffduration}, when the training data duration is greater than or equal to 4 hours, the improvement in model performance becomes markedly modest. From this, it can be concluded that fine-tuning and training speech recognition for dysarthric speakers, using pretrained models from large-scale datasets like WenetSpeech, requires approximately 5 hours of data. This finding aids in future dysarthric speech data collection, reducing database construction costs, especially when acquiring and annotating such data is challenging.

\section{Conclusion}
%我们发布了包含了134个小时构音障碍语音数据的中文构音障碍语音数据库（CDSD），数据采集自44个构音障碍者，其中包括5名儿童。该数据设置了不同的采集场景，保证音频高质量的同时还更贴合构音障碍者的日常生活场景。并且该数据库还包含了9个小时的构音障碍者说话时的视频数据，可以提供多模态的训练。该数据库是目前我们已知的最大的中文构音障碍数据库。通过对比实验和分析表明，CDSD可以为构音障碍语音识别模型提供很好的训练效果。
%我们相信，这个数据库是一个发展中文构音障碍语音识别的重要资源。我们希望通过数据库构建，提高中文构音障碍语音识别的可用性，通过这项工作为中文构音障碍人群提供实用且便利的语音识别技术。在未来的展望中，我们希望能继续拓展中文构音障碍数据库，以训练出构音障碍者通用的语音识别模型，为构音障碍者的社会交流与心理健康做贡献。

We have presented the Chinese Dysarthria Speech Database (CDSD), which includes 133 hours of data collected from 44 speakers.  
As we know, CDSD is the largest database of dysarthria in Chinese. The samples were collected in different scenarios and devices to ensure the high quality of the audio and to better match the daily life of dysarthric speech. The construction of the CDSD aims to improve the performance of DSR for Chinese, empowering practical and convenient DSR technology for individuals with dysarthria in China.
%Comparative experiments and analyses show that CDSD can provide good training results for speech recognition models of dysarthria.
\par
%We believe this database is essential for developing speech recognition for Chinese dysarthria. 

%通过对比实验分析表明，DSR有提高构音障碍者交流便利性的潜力。除此之外，仍需要更有效的特征和算法来解决不同构音障碍语音之间差异较大的问题。usability
 
%In the future, we plan to continue expanding the database of Chinese dysarthria to train an everyday speech recognition model for dysarthric people, contributing to their social communication and mental health.

%\iffalse
\section{Acknowledgements}
%Acknowledgement should only be included in the camera-ready version, not in the version submitted for review.
%The 5th page is reserved exclusively for \red{acknowledgements} and  references. No other content must appear on the 5th page. Appendices, if any, must be within the first 4 pages. The acknowledgments and references may start on an earlier page, if there is space.
Yan Wang and Mengyi Sun contributed equally to this research. We would like to express our gratitude to all the participants in our research. And we also extend our thanks to the Hangzhou Xiaoshan Noah Cerebral Palsy Service Center, the Angel House Rehabilitation and Educational Activity Center, Our Family China, and the Little Snail Family Support Center for People with Disabilities. This research was partially funded by 1) the National Natural Science Foundation of China ( 62276252, 62106256); 2) the Youth Innovation Promotion Association CAS. 

\bibliographystyle{IEEEtran}
\bibliography{reference}

\end{document}